\documentclass{article}

\usepackage{arxiv}
\usepackage{color}
\usepackage[utf8]{inputenc} % allow utf-8 input
\usepackage[T1]{fontenc}    % use 8-bit T1 fonts
\usepackage{booktabs}       % professional-quality tables
\usepackage{amsfonts}       % blackboard math symbols
\usepackage{nicefrac}       % compact symbols for 1/2, etc.
\usepackage{microtype}      % microtypography
\usepackage{lipsum}
\usepackage{graphicx}
\usepackage[hyphens,spaces,obeyspaces]{url}

\urldef\myurl\url{https://github.com/DP-3T/documents/blob/master/Security%20analysis/Privacy%20and%20Security%20Attacks%20on%20Digital%20Proximity%20Tracing%20Systems.pdf}

\urldef\myurll\url{https://eur-lex.europa.eu/legal-content/EN/TXT/?qid=1587141168991&uri=CELEX:52020XC0417(08)#ntr10-CI2020124EN.01000101-E0010}

\urldef\myurlll\url{https://eur-lex.europa.eu/legal-content/EN/TXT/?qid=1587141168991&uri=CELEX:52020XC0417(08)#ntr3-CI2020124EN.01000101-E0003}

\urldef\myurllll\url{https://github.com/DP-3T/documents/blob/master/DP3T%20White%20Paper.pdf}

\urldef\myurlllll\url{https://github.com/pepp-pt/pepp-pt-documentation}

\graphicspath{ {./images/} }

\title{Covid-19 and contact tracing apps: A review under the European legal framework}

\author{
  Teresa Scantamburlo \\
        European Centre for Living Technologies \\
        \texttt{teresa.scantamburlo@unive.it} \\
   \And
 Atia Cortés\\
        Barcelona Supercomputing Center\\
        \texttt{atia.cortes@bsc.es} \\
  \AND
  Pierre Dewitte \\
  Katholieke Universiteit Leuven\\
  \texttt{pierre.dewitte@kuleuven.be} \\
  \And
  Daphné Van Der Eycken \\
  Katholieke Universiteit Leuven\\
  \texttt{daphne.vandereycken@kuleuven.be} \\
  \And
  Valentina Billa \\
  Piccolotto\&Pierobon Law Firm  \\
  \texttt{vbilla@piccolottopierobon.com} \\
  \And
  Pieter Duysburgh \\
  Vjrie Universiteit Brussel\\
  \texttt{Pieter.Duysburgh@vub.be} \\
  \And
  Willemien Laenens\\
  Vjrie Universiteit Brussel\\
  \texttt{Willemien.Laenens@vub.be} \\
}

\begin{document}

\maketitle

\begin{abstract}
In this article, we would like to review the main technologies that have been proposed so far to fight the spread of the virus. Also, we would like to give an overview of the policy recommendations that some European organisations have put forward in these regards. Finally, we conclude with some considerations we would like to present to public attention and discussion.
\end{abstract}

% keywords can be removed

\section{Introduction}

Covid-19 is a global crisis which has mobilized huge forces in any affected country. Health care personnel are playing and essential role in treating the infection and taking care of people. Citizens are contributing with their behaviour as well by following measures for physical distancing and, in some cases, keeping themselves in isolation for at least 14 days. The computer science community is also doing its part. With the aim to contribute to a safe de-escalation of the confinement, some researchers and companies are working on the development of contact tracing apps. In general, the objective is to warn people that have recently been in contact with a person that is infectious in order to recommend quarantines to selected groups or individuals instead of extending the general physical distance.

Contact tracing is not a new in the treatment of epidemic disease. Implemented manually with human interviewers for monitoring the spread of Ebola virus disease (Swanson, 2018), it was also recommended by the World Health Organisation . But a recent study provided a mathematical model showing that a carefully designed app would render the whole process more effective (Ferretti et al, 2020). Now, the rapid growth of technical solutions for contact tracing solicited a debate on their practical effectiveness and, in particular, their medium and long-term effect on our society.

In this article, we would like to review the main technologies that have been proposed so far to fight the spread of the virus. Also, we would like to give an overview of the policy recommendations that some European organisations have put forward in these regards. Finally, we conclude with some considerations we would like to present to public attention and discussion.

\section{COVID-19 applications: a technical description}

Until now, the solutions proposed to governments to monitor the pandemic rely on the following technologies: Bluetooth, GPS and the network infrastructure. Another portion of COVID-19 applications employs self-reports and online forms. We briefly describe these approaches with a few examples in the following subsections.

\subsection{Self-assessment surveys}
Traditionally, contact tracing was handled through personal interviews between health professionals and the patients. This process required to generate a list of possible contacts during the past 2 weeks, although results could be very imprecise as they rely on the memory of the person interviewed. It was also difficult to measure if a contact was significant enough to be considered as a random contact or as a case to be analysed. Studies such as the one presented in Ferretti et al (2020) suggest that this technique is not enough to control a pandemic such as COVID-19 and the individuals at risk of transmission.

One of the first initiatives taken by several governments, research centres and other institutions was to create online survey forms or mobile apps to gather health information from self-assessments. In the case of mobile apps, this methodology is often combined with Bluetooth or GPS solutions. The common structure of these surveys starts by asking for some personal information: generally gender, age range and location (the level of detail depends on each survey and country, going from the name of the city, to the zip code or even the street name), but could also be extended to professional sector, level of income and others. Next, a series of questions are asked to reveal possible COVID-19 symptoms. All this information is collected and analysed in order to study the evolution of the pandemic and identify those areas where it was more present than in others, and ultimately to help governments and health authorities to take decisions.  Many of these surveys were initially published as Google forms, which presents some controversies from an ethical perspective: there was a lack of privacy statement, informed consent, the purpose of the investigation or the proper explanation of data storage and processing. Although most of them have tried to correct and improve those issues, there is still a lack of information of how all the previous data has been treated.

\subsection{Bluetooth}

Bluetooth is a wireless standard protocol for communication that is nowadays present in most mobile devices, such as cellphones, tablets, computers or smartwatches/smartbands. There are several types of Bluetooth protocols, although in the case of mobiles the Bluetooth Low Energy (BLE) protocol is the most commonly used due to its low price and consumption. We can find many applications, especially in the field of Internet of Things (IoT) that use BLE technology for communication between different types of devices. 
The BLE protocol is now being used as the base technology behind proximity contact tracing apps that are being developed as a way to monitor the evolution of the coronavirus pandemic. The characteristics of BLE against other technologies are, first the before-mentioned low power consumption, which allows contact tracing apps to run for hours without shortening mobiles' batteries too fast. Second, BLE works very well as a short-distance tracker and is especially beneficial in indoor environments.

The Pan-European Privacy-Preserving Proximity Tracing (PEPP-PT) project, led by the Fraunhofer Institute of Telecommunications, proposes a protocol based on BLE technology. 
Upon this proposal, a team led by Ecole Polytechnique Fédérale de Lausanne (EPFL), a research centre and university specializing in natural sciences and engineering, is working on a Decentralised Privacy-Preserving Proximity Tracing (DP-3T) protocol that aims to provide a low-cost solution to trace the pandemic while guaranteeing privacy of personal information. 

By decentralising the service, the group expects to increase the security of the app and the protection of personal data, avoiding any data breach that would lead to reidentification of end-users. At the same time, they seek for data minimisation by collecting only anonymous information about infected people and discarding any tracing information from non-infectious individuals. This is used by epidemiologists to build the network of contacted people with a possibility to have been infected, but no other personal or health information. At the same time, this protocol enhances user empowerment by giving them the choice to voluntarily share the information gathered by their mobile devices with health authorities. With these requirements, the DP-3T seeks to follow the General Data Protection Regulation (hereafter: GDPR) and the ePrivacy Directive, prioritising the limitation of data usage and storage, as it will be described in Section \ref{legal}.

Both protocols work with temporary or Ephemeral ID (EphID), which aims to be beneficial to preserve users' privacy since it changes frequently (the time base is customisable) and thus the identification of the real person is difficult. EphIDs will be exchanged among users that are found under a certain proximity distance. The team led by Carmela Troncoso, Assistant Professor at EPFL, has published the documentation of their protocol \footnote{See \myurllll} with an aim of obtaining feedback from the community. They are updating the document and notifying each new modification, as an effort of transparency and accountability. The PEPP-PT consortium has also provided a repository to access all their documentation \footnote{See \myurlllll}.

Both approaches aim to be used as standards by any national initiative that aims to develop a contact tracing app. It has been designed to support interoperability among different countries by following the European privacy and data protection regulations and principles. However, in the last few days, many partners – including KU Leuven – have withdrawn from the PEPP-PT consortium due to the lack of transparency around the protocol. The main criticism being the fact that, in case a user is diagnosed positive to COVID 19, the PEPP-PT’s protocol (NTK) preconizes the sending of the observed EphIDs to the backend server while DP-3T’s counterpart relies on the infected user’s own emitted EphIDs. Similarly, the computation of the risk happens on the backend server in the former case, while it is processed locally in the latter. 
DP-3T consortium describes the main conflicts that PEPP-PT’s protocol presents in terms of privacy and data protection\footnote{See \myurl}.

This is not the only approach that has been used for contact tracing apps. In Singapore, the government has developed BlueTrace, an open source protocol used in the TraceTogether PPPT app, which is also based on Bluetooth. The methodology is very similar: every time a mobile phone detects proximity with another device that is also using the app, it will that information store locally and share it with the health authorities. The main difference with the European initiatives is the encryption method but both seek for being as less invasive as possible.
Apple and Google have also started a joint initiative to design their own protocol which is based on DP-3T, which will be directly installed as a feature of their respective operating systems iOS and Android.

\subsection{Global Positioning System}

GPS (Global Positioning System) is a satellite-based radionavigation system which is operated by the United States Space Forces. Usually, modern mobile phones incorporate a special type of GPS, i.e. the Assisted Global Positioning System (AGPS), which improves the performance of a stand-alone GPS by leveraging the wireless network of mobile’s operator. GPS uses radio waves between satellites and a receiver inside the mobile phone to provide location and time information to any application that needs to use it (like Google Maps). Geolocalisation services are many and range from emergency calls to tracking, sharing location data with friends and location-based ads. Note that since the AGPS uses the cellular network, the precision of the geolocation data depends on the strength of the signal between the phone and the cell tower. Although new GPS chips promise to pinpoint a device’s accuracy to within 30 centimetres (Kastrenakes, 2017), today's solutions offer typically a range of three to five meters and in certain environments (e.g. underground locations) they may lack precision. For this reason, inside buildings like airports or car parks, indoor positioning systems, based for instance on Bluetooth or a wireless local area network, are preferred. An example of an app using GPS is SOS Italia, where location data are collected only after users' authorization, \textit{i.e.} activation is not automatic. Also, SOS Italia relies on a digital platform accessible only to authorized persons, responsible for collecting and organizing the information entered in the app directly from citizens, starting with name, surname and personal data on residence and domicile, accompanied by their tax code (Rusconi, 2020). 

In general, COVID-19 apps using GPS turn out to be problematic for its invasive tracking and several governments have already stopped some projects. For example, Israel’s High Court of Justice put a stop to a COVID-19 app using citizens’ smartphones to track the locations of infected people (Bandel, 2020). A list of COVID-19 using geolocation data can be found on the website of the European Global Navigation Satellite Systems Agency .  

\subsection{Network infrastructure}

Network infrastructure can also be used to track the position of a mobile phone. The system uses the signal that a mobile phone exchanges with the closest cell tower of the network. The primary difference between the network-based technique and the GPS is the way in which they gather location data. GPS uses satellites that orbit around the Earth to triangulate a user's location, whereas network-based techniques use the service provider's network infrastructure. But most of today's smartphones combine the two technologies (such as the AGPS). The network infrastructure allowed telecommunication companies to collect data to track the virus. For example: Vodafone stated that it produced “an aggregated and anonymous heat map for the Lombardy region in Italy to help the authorities to better understand population movements in order to help thwart the spread of COVID-19.”  In general, the network infrastructure alone provides poorer results as compared to AGPS and Bluetooth as knowing that a number of people in a relatively small area, say 50 meters, would not be very informative in determining possible infections.

\section{Legal considerations} \label{legal}

While the use of mobile apps can considerably streamline contact tracing – otherwise based on individual interviews with infected patients –, it also raises numerous legal challenges when it comes to guaranteeing individuals’ fundamental rights to privacy and data protection. The following paragraphs aim at highlighting the most pressing issues surrounding the development, the rolling out and dismantling of digital technologies assisting in monitoring and containing the COVID-19 pandemic.

As a preliminary remark, it is worth highlighting that COVID-19 mobile apps can meet different needs. The European Commission acknowledged\footnote{See the recommendations released by the European Commission in April 2020 and available online \url{https://ec.europa.eu/info/sites/info/files/recommendation_on_apps_for_contact_tracing_4.pdf}} that they generally fulfil four different functions: 
\begin{enumerate}
    \item informing and advising citizens about the virus and the general measures adopted by Member States;
    \item providing self-assessment questionnaires, symptom checkers or telemedicine functionalities;
    \item performing contact tracing to inform people who have been in proximity to an infected patient;
    \item providing a communication forum between patient and doctors in situation of self-isolation or where further diagnosis and treatment advice is provided (increased use of telemedicine);
\end{enumerate}{}
While only the second and third categories of apps are likely to raises data protection issues, the present contribution will focus on apps supporting contact tracing features.

The technical solutions that will be adopted by governments will also have to take into account many legal, societal and ethical implications. These issues are strictly connected. Indeed, measures must comply with international human rights law and cannot consist in disproportionate and unnecessary actions. In accordance with Art. 15 of the European Convention on Human Rights, in time of emergency, measures that may infringe on rights must be taken \textit{“to the extent strictly required by the exigencies of the situation, provided that such measures are not inconsistent with its other obligations under international law”}. A similar provision can be found in Art. 52, par. 1, of the EU Charter of Fundamental Rights: any limitation on the exercise of the rights and freedoms recognised by this Charter must be provided for by law, subject to the principle of proportionality, made only if necessary and must not excessively affect the fundamental right to a private life. 
In addition, it is necessary to consider the ePrivacy Directive and the General Data Protection Regulation (GDPR), according to which location data is personal data, and therefore is subject to high levels of protection even when processed by public authorities or private companies \footnote{See the statement on the processing of personal data in the context of the COVID-19 outbreak published by the European Data Protection Board \url{https://edpb.europa.eu/our-work-tools/our-documents/other/statement-processing-personal-data-context-covid-19-outbreak_en}.}.

\subsection{The European legal framework}

First, in order to verify whether such apps fall within the material scope of application of the GDPR –, it is necessary to check whether they involve the ‘processing’ (Art. 4(2) GDPR) of ‘personal data’, \textit{i.e.} any information relating to an identifiable natural person (Art. 4(1) GDPR). Regardless of the technology used to perform contact tracing, the general consensus is that they do. While it is rather obvious for contact tracing based on geolocation data – the privacy-invasive and repurposing potential of which is well-documented – the same goes for BLE-based solutions, deemed as the most privacy-preserving alternatives. As highlighted in the privacy analysis of the DP-3T protocol, even decentralised options based on the sharing of emitted EphIDs are vulnerable to re-identification attacks and, therefore, would warrant the qualification of the data processed as ‘personal’\footnote{See the documentation online: \myurl}. Without delving into the intricacies of the ‘identifiability’ threshold under Art. 4(1) GDPR, one can reasonably assume that contact tracing apps will fall under the scope of application of the GDPR and, as such, will need to abide by the various principles and rules prescribed therein. 

Second, it is crucial to identify and adequately qualify the actors involved in the processing operations, as this will determine the allocation of responsibility, liability and accountability under the Regulation. In other words, who will be responsible, and will be held liable, for compliance with EU data protection law. Under the GDPR, that entity is the ‘controller’, \textit{i.e.} the one that determines the ‘purposes’ and the ‘means’ of the processing activities. While various options can be considered – involving both public and private actors –, the European Commission recommends a model where such responsibility would fall to national health authorities, or the entity carrying out tasks of public interest in the field of health. This, underlines the Commission and the European Data Protection Board (hereafter: EDPB), is essential to foster public trust and guarantee sufficient adoption \footnote{Communication of the European Commission, point 3.1 (\url{https://eur-lex.europa.eu/legal-content/EN/TXT/?qid=1587141168991&uri=CELEX:52020XC0417(08)}) and the EDPB' Opinion 4/2020, para. 25 (\url{https://edpb.europa.eu/sites/edpb/files/files/file1/edpb_guidelines_20200420_contact_tracing_covid_with_annex_en.pdf})}.

Besides the applicability of the GDPR, it is to be noted that the data produced via the smart devices are also protected as ‘location data’ under the ePrivacy Directive . Additionally, Article 5(3) ePrivacy Directive applies to every entity that places on or reads information from smart devices. It applies without regard to the nature of the entity and includes public as well as private actors. Concretely, Article 5(3) of the ePrivacy Directive prescribes that storing information on a user’s device or gaining access to information already stored is allowed only with consent of the user or if the storage and/or access is strictly necessary for the app installed or activated by the user.

Operators of mobile apps offering contact tracing functionalities will need to follow a security and data protection by design approach. The following describes the principles and rules laid down in the GDPR that should be taken into account when developing, designing, selecting and using applications that are based on the processing of personal data. 

\subsection{Lawfulness}
In relation to the installation of the app and the storing and accessing of information on the user’s device, ePrivacy Directive requires either \textit{(i)} the user’s freely given, specific, informed and unambiguous consent or \textit{(ii)} to justify that the storage and access is strictly necessary to ensure the proper functioning of a service explicitly requested by the user. While the latter option could cover situations where the user wishes to be informed of a contact with an infected person, the upload of his/her EphIDs to the backend server will require the user’s consent since it is not strictly necessary for the functioning of the app. Given that the uploaded information will likely be considered as a ‘special category of personal data’ (\textit{i.e.} ‘personal data revealing racial or ethnic origin, political opinions, religious or philosophical beliefs, or trade union membership, and the processing of genetic data, biometric data for the purpose of uniquely identifying a natural person, data concerning health or data concerning a natural person's sex life or sexual orientation’), that consent must also be made explicit, that is, expressed through a clear affirmative action (Art. 9(2)a GDPR).

When it comes to the processing of personal data by national health authorities, both the EDPB and the European Commission favour the use of Art. 6(1)c and e, together with Art. 9(2)g and i GDPR, as adequate lawful grounds. In other words, pre-existing or specifically-enacted national legislation laying down the ins and outs of the processing operations happening within the context of the mobile app as well as the guarantees deployed to safeguard the rights and freedoms of data subjects (Art. 6(3) GDPR). This option presents the noteworthy advantage of clearly delineating, in a legally binding instrument, the controller(s), the potential recipients, the processing operations and the purposes. From a legal perspective, this also opens up the possibility to scrutinize the said national laws through the lenses of the three-step test applied to verify whether an interference with individuals’ fundamental rights is justified (\textit{i.e.} legality, necessity, proportionality). As highlighted by all regulatory actors, the installation and use of the app should only happen on a voluntary basis regardless of the lawful ground used to justify the processing. In the same vein, refusal to install the app should not lead to adverse consequences for individuals.

\subsection{Transparency}

Both the GDPR and the ePrivacy Directive requires controllers to be transparent about their processing activities. The operator of the mobile app will therefore need to make the information listed in Art. 13 GDPR – and, if relevant, Art. 14 GDPR – readily available to the data subject in a concise, transparent, intelligible and easily accessible form, using clear and plain language. This should include, for instance, the identity and contact details of the controller, the purposes and lawful ground of the processing, the recipients of personal data if any, the retention period and the existence of the multiple prerogatives granted to data subjects such as the right to access and erasure. Armed with that information, individuals should be able to make informed choices as to the installation and use of the contact tracing app.

Also, contact tracing apps should be developed in a transparent and verifiable manner, paving the way for public scrutiny. Open source code, external audits and publicly available Data Protection Impact Assessments (DPIA) are, in that sense, examples of best practices fostering citizens’ engagement.

\subsection{Purpose limitation}

Controllers must also comply with the principle of purpose limitation, which requires personal data to be \textit{(i)} collected for explicit, specified and legitimate purposes, \textit{i.e.} purpose specification, and \textit{(ii)} not further processed in a manner that is incompatible with those purposes, \textit{i.e.} compatibility assessment (see Art. 5(1)b GDPR). Applied in the present context, this means that the operators of the mobile app must explain, in clear and plain language, the reason(s) why each category of personal data is processed (\textit{e.g.}, in the context of BLE-based contact tracing apps, the emitted EphIDs of a patient diagnosed positive with COVID-19 are sent to the backend server in order for the system to inform, via push notification, the other users of the app that they have been in contact with an infected person). It is crucial that the purpose of the processing of personal data should be limited to fighting the current sanitary crisis, and nothing else. In the same vein, controllers must refrain from using the collected personal data for purposes that are incompatible with contact tracing (\textit{e.g.}, in the context of symptom checker apps, using the answers to the questions to target individuals with personalised ads). One way to ensure compliance with that principle is to only collect personal data the repurposing potential of which is limited, such as ephemeral identifiers. Another is to avoid the bundling of functionalities within the same app (\textit{e.g.}, a single app providing general information, symptom checker features and contact tracing) or, should that be the case, to clearly distinguish those functionalities and grant users granular control over which of them he or she wishes to opt-in to. 

\subsection{Data minimisation}

It is crucial to ensure compliance with the data minimisation principle, which requires controllers to only collect and further process personal data that is necessary in relation to the purposes that have been specified (Art. 5(1)c GDPR). In the present context, it raises the issue as to whether or not a given piece of information is strictly necessary to ensure the proper functioning of contact tracing or, to put it differently, whether there is a less privacy-invasive way to yield identical results. In its recommendation of 8 April, the European Commission\footnote{See \url{https://ec.europa.eu/info/sites/info/files/recommendation_on_apps_for_contact_tracing_4.pdf}.} already stated that the use of geolocation and/or movement data should be avoided, as this information is not strictly necessary to discover events such as the contact with an infected patient. Alternatives based on BLE are, in the institution's view, both more efficient and less privacy-invasive. This has been later acknowledged by the EDPB\footnote{See Opinion 4/2020, para. 27.} and included as an essential requirement in the Common EU Toolbox for Member States developed by the eHealth Network (see Annex 1, Id. CS-03 and SG-03). In its latest communication of 17 April, the European Commission\footnote{See \myurll} has also emphasised that it is not necessary to store the exact time of the contact nor the place, since knowledge of the day alone would be sufficient to calculate the risk factor. Same goes for any type of unnecessary metadata that is not specific to a contact between people and its duration, according to the Chaos Computer Club\footnote{See requirement 6 in \url{https://www.ccc.de/en/updates/2020/contact-tracing-requirements }}. 

\subsection{Storage limitation}

Controllers will need to comply with the storage limitation principle, which requires to tailor the retention period according to the purposes of the processing (Art. 5(1)d GDPR). In other words – and as acknowledged by the European Commission\footnote{See \myurll and also EDPB Opinion 4/2020, para. 3} –, proximity data should be deleted as soon as they are no longer necessary for the purpose of alerting individuals. In the case of a decentralised, BLE-based solution, this means deleting the ephemeral identifiers collected by the user during his or her encounters after the incubation period. Same goes for personal data potentially stored on the backend server used to let other users know about contacts with infected persons.

\subsection{Accuracy, integrity and confidentiality}

Accuracy, integrity and confidentiality are key principles when it comes to the development of contact tracing apps (Art. 5(1)d and f GDPR). Guidelines and suggestions as to the appropriate safeguards to substantiate those rather abstract notions have been issued at various levels, by both public and private actors. Among the most relevant ones is the need to ensure that personal data are processed – as much as technically feasible –, on the user’s device rather than by a centralised platform. The consortium behind the DP-3T protocol advocates, for instance, for the computation of the risk factor to happen locally. Additionally, the European Commission advocates for the use of state-of-the art cryptographic techniques for generating, storing and transmitting proximity data. In the case of BLE-based proximity tracing, the institution also recommends using temporary user IDs that change regularly rather than fixed identifiers linked to the user’s device. All in all, it should not be possible for the receiving user to re-identify the individuals using the emitted ephemeral identifiers, and vice versa. While dynamic identifiers remains prone to some re-identification attacks, the architecture of the system should tend toward proper anonymisation, highlights the Chaos Computer Club (see requirement 7).

\subsection{The guidelines of the European Data Protection Board}

Recently, on 21 April 2020, the EDPB adopted the “Guidelines 04/2020 on the use of location data and contact tracing tools in the context of the COVID-19 outbreak”. 
The Guidelines concern the use of location data and contact tracing tools for two specific purposes:
\begin{itemize}
    \item to model the spread of the virus so as to assess the overall effectiveness of confinement measures; 
    \item to notify individuals of the fact that they have been in close proximity of someone who is eventually confirmed to be a carrier of the virus.
\end{itemize}{}

The EDPB refers to GDPR and Directive 2002/58/EC (the “ePrivacy Directive”) and recalls the general principles of effectiveness, necessity, and proportionality that must be taken into account for any measure involving the processing of personal data adopted by Member States or EU institutions in the fight against COVID-19.

Where possible, the apps should use anonymised data (using techniques that ensure no linkages between the data and an identified or identifiable natural person can be made) instead of pseudonymised data (that is in the scope of GDPR). The Guidelines also clarify the differences between anonymised and pseudonymised data, explaining the three criteria in accordance with which evaluating the robustness of anonymisation: 
\begin{enumerate}
    \item singling-out (isolating an individual in a larger group based on the data); 
    \item linkability (linking together two records concerning the same individual); 
    \item inference (deducing, with significant probability, unknown information about an individual).
\end{enumerate}{}

In order to increase the trust of the users, the EDPB suggests to make the anonymisation methodology transparent. 

The Guidelines carry out a complete legal analysis of the contact tracing applications, where the above-mentioned data quality principles can be found.   

Moreover, if, on one hand, the effectiveness of these applications is related to the implementation of other complementary measures, on the other hand, the EDPB points out that such apps cannot replace, but only support, manual contact tracing performed by qualified public health personnel. Also, it claims that the procedures including use of algorithms implemented by the contact tracing apps should work under the strict supervision of qualified personnel. In addition, states the EDPB, algorithms must be auditable and should be regularly reviewed by independent experts. 

Another relevant issue strictly related to a human evaluation is the storage of data, connected for example to epidemiological considerations about the incubation period of the virus; in any case, as the EDPB highlights, the data should be erased or anonymised after the end of the crisis.

\section{Discussion}

The common weakness of all these solutions is that they rely on the individual adoption of the different methodologies presented. Will it ever be possible to attain a sufficient uptake of the tracing app to guarantee an impact on the spread of the COVID-19 virus? In the case of COVID-19, there are no previous data sets nor research focused on this problem, since it is the first time that the humanity is facing it. Also, the effectiveness of tracing apps depends on the number of people using them.
For example, a group of researchers at the university of Oxford reported that 80\% of UK users of smartphone devices (56\% of the population) would need to install and use the app in order to be effective (Hinch et al., 2020).
These numbers are far from being representative of the whole population of a country, taking into account that there are specific groups that will be excluded from the sampled data: older adults and young kids, but also people with low incomes or no access to digital resources. A survey carried by Digimeter \footnote{\url{https://www.imec-int.com/en/imec-digimeter-2019}} showed that 27\% of Flemish people aged 65 years and older do not even own a smartphone. Even when sufficient trust can be created for a widespread adoption of the app or if the installation of the app will be obligatory, we still need to keep in mind that a substantial part of population does not own a smartphone. To provide technical robustness, the possible outcomes of such apps and surveys require of vast amount of new information, which should be verified and validated to provide trusted results. In addition, the solutions provided need to be socially robust as well, and thus it is important to not leave aside the vulnerable groups mentioned before and increase the already existing digital gap.

In its communication of 17 April 2020, the European Commission\footnote{See \myurlll} has expressed a strong preference for BLE-based solutions as it is more precise than geolocation data for the metering of proximity and avoids the collection of unnecessary movement data. In the recent “Guidelines 04/2020 on the use of location data and contact tracing tools in the context of the COVID-19 outbreak”, the EDPB states that “it seems preferable to develop a common European approach in response to the current crisis, or at least put in place an interoperable framework”. However, However, while Europe made its preference for BLE-based solutions clear, there are still unknown aspects surrounding national options which may not guarantee a full commitment to European values and its legal framework.

Many tech solutions proposed so far try to follow the principle of privacy by design, \textit{i.e.} they try to implement technical and organisational measures to prevent privacy breaches and to be compliant with the GDPR. For example, the PEPP-PT project claimed that the designed system is based on security and privacy safeguards, such as data minimization and graceful dismantling (the system will organically dismantle itself after the end of the epidemic). Also, the proponents published a white paper to document and describe to the large public the properties of the system with a view to get feedback which can improve privacy aspects. While these and similar initiatives are fundamental to guarantee a safe introduction of tracing apps into society in these unique and exceptional conditions, we argue that privacy by design is not enough. Technical and organisational measures cannot substitute the human effort needed to build a culture of trust among citizens and between citizens and institutions. Safety is not only a matter of a technical or an organisational provision. It is also a condition which reflects a relationship between the ones who expect protection and those who take the responsibility to fulfil this expectation. To ensure the success of technical interventions we need to sustain a culture of trust among people promoting, for instance, a constructive and inclusive dialogue, a non-violent language, transparent communication and a responsible data management.

%Focusing just of tech solutions can also move our attention away from connected issues: the deficiency of the EU public healthcare system (many hospitals were closed, the public service handed over to private organisations, there was no prevention plan for pandemic) and environmental crisis \footnote{see also here: \url{https://theconversation.com/coronavirus-is-a-wake-up-call-our-war-with-the-environment-is-leading-to-pandemics-135023}}. 

% education / cohabit with tracing apps and the future of society 

It is in times like this that people will accept more easily some conditions that might have an immediate social impact, affecting our privacy and autonomy, seeking for a quick recovery of our past routines and lifestyles, the so-called "normality". A recent survey in Flanders (Belgium) showed that 1 in 2 persons is currently willing to install a tracing app voluntarily \footnote{See \url{https://www.data-en-maatschappij.ai/en/news/survey-corona-app}}. The spread and acceptance of contact tracing apps and other methods of surveillance are justified under the premise that it will be beneficial for our health status and healthcare systems, but is it helpful? Is it necessary to choose one side in this trade-off between privacy and health in order to control this pandemic? As Yuval Harari, an Insraeli historian philosopher and author of ‘Sapiens’ and ‘Homo Deus’, says, we should be able to enjoy both privacy and health, but for this we also need to empower citizens \footnote{See \url{https://www.ft.com/content/19d90308-6858-11ea-a3c9-1fe6fedcca75}} by offering them trustful sources of verified information.
In the “Guidelines 04/2020 on the use of location data and contact tracing tools in the context of the COVID-19 outbreak”, the EDPB underlines that \textit{“one should not have to choose between an efficient response to the current crisis and the protection of our fundamental rights: we can achieve both, and moreover data protection principles can play a very important role in the fight against the virus”}. It is clear that there is a need for guidelines and regulators to protect our fundamental human rights and prevent that our society moves from an emergence situation to a situation of exception. It is also the moment for Europe to work in Union and promote trust in a common initiative that ensures data governance, otherwise European laws and regulations on data privacy will not be guaranteed.

\section*{Acknowledgement}
Dr. Atia Cort\'es and Dr. Teresa Scantamburlo are partially supported by the project A European AI On Demand Platform and Ecosystem (AI4EU) H2020-ICT-26 \#825619. The views expressed in this paper are not necessarily those of the consortium AI4EU.

The authors would like to thank Dr. Jan de Bruyne, from the Katholic University of Leuven and the Centre Data \& Society for making this collaboration possible.

\section*{References}

Swanson KC, Altare C, Wesseh CS (2018). “Contact tracing performance during the Ebola epidemic in Liberia”, 2014–2015. PLoS Negl Trop Dis, 12

Kastrenakes J (2017). “GPS will be accurate within one foot in some phones next year”, The Verge, 25 September, \url{https://www.theverge.com/circuitbreaker/2017/9/25/16362296/gps-accuracy-improving-one-foot-broadcom}. Accessed 18 April 2020

Rusconi G. (2020). “Pronta l’app per il monitoraggio dell’epidemia. Ecco come potrebbe funzionare”. Il sole 24 ore, 26 Marzo 2020 \url{https://www.ilsole24ore.com/art/pronta-l-app-il-monitoraggio-dell-epidemia-ecco-come-potrebbe-funzionare-ADl0m7F}  Accessed 17 April 2020

Bandel N (2020). “Israel's Top Court: No Shin Bet Tracking of Coronavirus Patients Without Knesset Oversight”, Haaretz, 19 March 2020 \url{https://www.haaretz.com/israel-news/.premium-israel-s-top-court-no-shin-bet-tracking-of-coronavirus-patients-without-knesset-ove-1.8690253} Accessed 18 April 2020

Hinch R et al (2020), “Effective Configurations of a Digital Contact Tracing App: A report to NHSX”, report available on Github: \myurl

Luca Ferretti, Chris Wymant, Michelle Kendall, Lele Zhao, Anel Nurtay, Lucie Abeler-Dörner, Michael Parker, David Bonsall, Christophe Fraser (2020). “Quantifying SARS-CoV-2 transmission suggests epidemic control with digital contact tracing”, Science, 31 Mar 2020

%\bibliographystyle{abbrv}  

\end{document}